\documentclass[aps,physrev,preprint,superscriptaddress,floatfix]{revtex4-2}
\usepackage{amsmath,amssymb}
\usepackage{graphicx}
\usepackage{bm}

\begin{document}

\title{Mechanisms of turnkey soliton generation in self-injection-locked photonic-crystal ring resonators}

\author{Naoya Kuse}
\email{kuse.naoya@tokushima-u.ac.jp}
\affiliation{Institute of Post-LED Photonics, Tokushima University, 2-1, Minami-Josanjima, Tokushima, Tokushima 770-8506, Japan}
\affiliation{Institute of Photonics and Human Health Frontier, Tokushima University, 2-24, Shinkura-cho, Tokushima 770-8501, Japan}

\date{\today}

\begin{abstract}
Turnkey generation of dissipative Kerr solitons in photonic-crystal ring resonators relies on the interplay between engineered mode splitting, self-injection locking, and modulation instability, yet the mechanism governing soliton initiation and number selection remains incompletely understood. We investigate this interplay using coupled Lugiato–Lefever equations with a self-consistent nonlinear self-injection-locking condition. Analysis of continuous-wave steady states and their locking stability reveals operating points accessible through feedback-induced transitions between split resonances. Evaluating modulation-instability gain at these points identifies conditions for comb initiation and links the dominant unstable sideband modes to soliton-number selectivity. Direct time-domain simulations corroborate the analysis, showing that appropriately chosen mode splitting enables highly repeatable single-soliton generation without external laser-frequency scans. The final soliton number is nevertheless shaped by subsequent nonlinear evolution and feedback-induced detuning changes, rather than by the initial instability alone. We also show that larger dispersion suppresses higher-order unstable modes and favors lower soliton numbers.
\end{abstract}

\maketitle


\section{Introduction}

Dissipative Kerr solitons (DKSs) \cite{Herr2014,Kippenberg2018} in optical microresonators provide a compact source of coherent optical frequency combs with applications in optical communications \cite{MarinPalomo2017,Corcoran2025}, precision metrology \cite{Newman2019, wu2025vernier}, microwave generation \cite{sun2024integrated, kudelin2024photonic}, and optical ranging \cite{Riemensberger2020, Sun2023}. Early demonstrations of DKS microcombs typically required high-power, narrow-linewidth tunable lasers and bulky external optical components \cite{Brasch2016}. Advances in the fabrication of low-loss, high-quality-factor microresonators have substantially reduced the required pump power \cite{Liu2021,Ji2024}. 
However, reliable soliton generation remains challenging because the laser--cavity detuning must be carefully controlled, particularly during soliton initiation, when changes in intracavity power induce thermal shifts of the cavity resonance~\cite{Zhou2026}.

\begin{figure}[b]
\centering
\includegraphics[width=\linewidth]{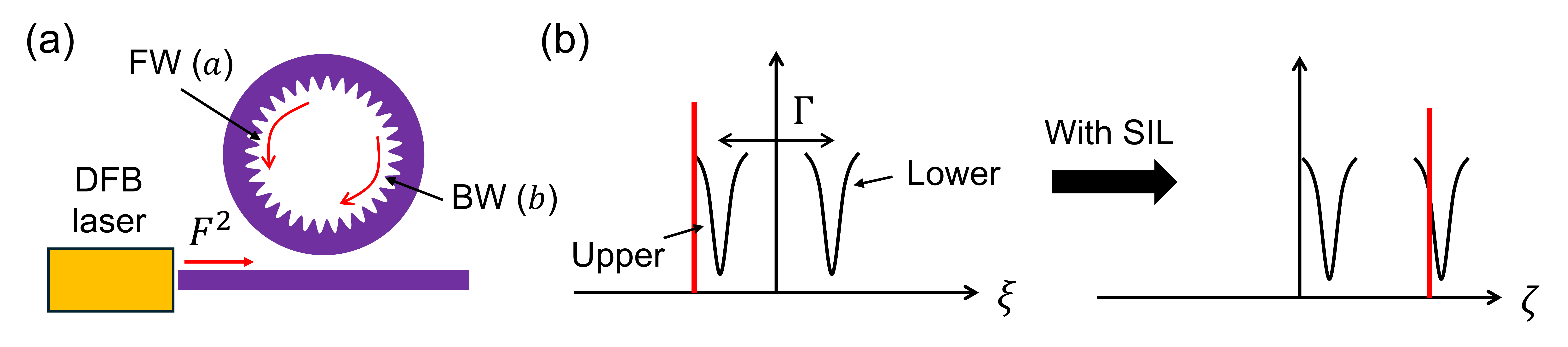}
\caption{(a) Illustration of a PhCR under SIL. (b) Schematic of the split resonances without (left) and with (right) nonlinear SIL. As the pump laser approaches the upper resonance, optical feedback induces SIL, while the Kerr nonlinearity shifts the resonance frequencies. The resulting frequency pulling allows the pump laser to transition from the upper resonance to the upper CW branch of the lower resonance.}
\label{fig:schematic}
\end{figure}

Self-injection locking (SIL) has emerged as an effective approach to overcoming this challenge. In SIL, light backscattered from the microresonator is fed back into the pump laser, as shown in Fig.~\ref{fig:schematic}(a), allowing the laser frequency to be stabilized through optical feedback~\cite{Kondratiev2023}. The Kerr nonlinearity has been shown to significantly modify the SIL tuning characteristics, enabling the laser to access detunings where soliton states can exist~\cite{Voloshin2021}. Combined with low-loss integrated microresonators, SIL has enabled turnkey DKS generation without requiring rapid laser-frequency scanning or complex active detuning control~\cite{Stern2018,Shen2020,Voloshin2021,Xiang2021}. 
However, single- and multi-soliton states can coexist under SIL, making reliable single-soliton selection challenging. To investigate soliton-number selection under SIL, Deshmukh et al.~\cite{Deshmukh2026} analyzed the bifurcation structure of soliton states in microresonators with weak backscattering and identified parameter regions where only the single-soliton branch is stable among the soliton branches. They further demonstrated access to single solitons through prescribed scans of the free-running laser detuning and feedback phase.

Photonic-crystal ring resonators (PhCRs) provide an additional degree of freedom for controlling soliton generation through engineered resonance splitting~\cite{Lu2014,lu2020universal,Yu2021,Lucas2023,Zang2025}. In a PhCR, periodic modulation of the resonator boundary induces coherent coupling between the forward-propagating (FW) and backward-propagating (BW) fields (Fig.~\ref{fig:schematic}(a)), resulting in controllable mode splitting at selected resonances (Fig.~\ref{fig:schematic}(b)). In our previous work, we showed that splitting the pumped resonance modifies the continuous-wave (CW) resonance structure and modulation-instability (MI) conditions, enabling soliton formation through dynamics that differ from those in conventional ring resonators~\cite{Sakaue2026}. Furthermore, the engineered FW--BW coupling provides a controllable optical feedback signal for SIL, in contrast to conventional microresonators that rely on random backscattering. Using this synthetic reflection, Ulanov et al.~\cite{Ulanov2024} demonstrated deterministic single-soliton generation in a self-injection-locked PhCR.
Their analysis identified two important roles of engineered mode splitting in turnkey soliton generation. First, stronger FW--BW coupling enhances the reflected field and extends the detuning range accessible through SIL, increasing its overlap with the DKS existence range. 
Second, appropriate mode splitting and pump power allow MI to occur with the maximum gain at the first sideband pair ($m=1$), favoring single-soliton formation. Based on these principles, they derived approximate analytical criteria for MI initiation and single-soliton selection. However, the overlap between the SIL-accessible detuning range and the DKS existence range alone does not determine which CW states are reached during fixed-parameter turnkey operation or whether comb generation is initiated. Furthermore, the analytical criterion favoring single-soliton formation assumes approximately equal FW and BW intracavity pump powers, limiting its applicability to more general coupling and pump-power conditions. For example, at a normalized pump power of $F^2=4$, the approximate criteria for MI initiation and $m=1$ selection cannot be satisfied simultaneously for any mode-splitting strength $\Gamma$. 
These limitations highlight the need to clarify how the nonlinear SIL tuning curve and CW stability determine the operating points accessible in PhCRs and how the MI characteristics at these points influence comb initiation and soliton-number selectivity.

In this work, we investigate the physical mechanisms governing turnkey soliton initiation and number selection in self-injection-locked photonic-crystal ring resonators. Although engineered mode splitting can extend the detuning range accessible through SIL, access to a soliton-supporting detuning does not by itself establish how comb generation is initiated or which soliton state ultimately emerges. Using coupled Lugiato–Lefever equations with a self-consistent nonlinear SIL condition, we relate the accessibility and stability of continuous-wave states to their modulation-instability characteristics and subsequent nonlinear evolution. We show that comb initiation and final soliton-number selection are related but distinct processes, providing a physical basis for understanding how engineered mode splitting enables reliable single-soliton generation.

\section{Theory}

To describe soliton microcomb generation in a PhCR under SIL, we consider coupled LLEs for the FW and BW intracavity fields \cite{Coen2013,Lucas2023,Sakaue2026}. Periodic modulation of the PhCR boundary induces coherent coupling between the counter-propagating waves and produces a controllable splitting of the pumped resonance. In contrast to a resonator driven by a laser with an externally prescribed frequency, the actual pump frequency in an SIL system is modified by optical feedback from the resonator, as illustrated in Fig.~\ref{fig:schematic}(b). We therefore describe the cavity dynamics and the laser-frequency pulling self-consistently. In the frequency domain, the normalized coupled LLEs are written as
\begin{equation}
\frac{\partial \tilde a_\mu}{\partial \tau}
=
-\left(
1+i\zeta
-i\frac{d_2}{2}\mu^2
\right)\tilde a_\mu
+i\mathcal{F}\!\left(|a|^2a\right)_\mu
+2i\left(\sum_{\mu'}|\tilde b_{\mu'}|^2\right)\tilde a_\mu
+i\frac{\Gamma}{2}\delta_{\mu 0}\tilde b_\mu
+F\delta_{\mu 0}
\label{eq:phcr_sil_lle_a}
\end{equation}
\begin{equation}
\frac{\partial \tilde b_\mu}{\partial \tau}
=
-\left(
1+i\zeta
-i\frac{d_2}{2}\mu^2
\right)\tilde b_\mu
+i\mathcal{F}\!\left(|b|^2b\right)_\mu
+2i\left(\sum_{\mu'}|\tilde a_{\mu'}|^2\right)\tilde b_\mu
+i\frac{\Gamma}{2}\delta_{\mu 0}\tilde a_\mu
\label{eq:phcr_sil_lle_b}
\end{equation}

Here, $\tilde a_\mu$ and $\tilde b_\mu$ are the normalized complex modal amplitudes of the FW and BW fields, respectively, with the mode index $\mu$ measured relative to the pumped mode ($\mu=0$). The corresponding angular-domain fields are $a(\theta,\tau)=\sum_\mu\tilde a_\mu e^{i\mu\theta}$ and $b(\theta,\tau)=\sum_\mu\tilde b_\mu e^{i\mu\theta}$. The normalized slow time and second-order dispersion coefficient are $\tau=\kappa t/2$ and $d_2=2D_2/\kappa$, where $\kappa$ is the total cavity energy decay rate. The loss rate is consequently normalized to unity. The terms proportional to $\zeta$ and $d_2\mu^2/2$ represent laser--cavity detuning and second-order dispersion, respectively. The Fourier operator $\mathcal{F}(\cdot)_\mu$ extracts the $\mu$-th spectral component of the Kerr self-phase-modulation term, while the terms proportional to $2\sum_{\mu'}|\tilde b_{\mu'}|^2$ and $2\sum_{\mu'}|\tilde a_{\mu'}|^2$ describe cross-phase modulation arising from the total power in the counter-propagating field. The parameter $\Gamma$ denotes the normalized frequency splitting of the pumped resonance, so that $\Gamma/2$ is the FW--BW coupling amplitude. The Kronecker delta $\delta_{\mu0}$ restricts this linear coupling to the pumped mode. The external driving term $F\delta_{\mu0}$ is applied only to the FW field. The BW field is generated through intracavity coupling. The pump phase is chosen such that $F$ is real.

The effective laser--cavity detuning $\zeta$ is not an independent control parameter in the SIL configuration. We distinguish the free-running laser detuning, $\xi=2(\omega_0-\omega_{\mathrm{LC}})/\kappa$, from the effective detuning, $\zeta=2(\omega_0-\omega_{\mathrm{eff}})/\kappa$, where $\omega_0$ is the unsplit reference resonance frequency, $\omega_{\mathrm{LC}}$ is the free-running laser frequency, and $\omega_{\mathrm{eff}}$ is the oscillation frequency in the presence of optical feedback. 
The field returned to the laser is determined by the BW component at the pumped mode. 
Assuming that the pump laser gain relaxation time ($\approx$ 0.1 ns) is much shorter than the microresonator response time (> 10 ns), we adiabatically eliminate the laser-amplitude dynamics \cite{Shen2020,Wang2025} and impose the stationary phase-locking condition \cite{Shen2020,Voloshin2021}. We further neglect explicit feedback-delay dynamics and absorb the approximately constant propagation phase, together with the laser linewidth-enhancement contribution, into the effective feedback phase \cite{Shen2020,Voloshin2021,Wang2025}. The resulting quasi-static SIL condition is
\begin{equation}
\zeta
=
\xi
+
\frac{K_0\Gamma}{2}
\operatorname{Im}
\left[
e^{i\bar\psi}
\frac{\tilde b_0}{i(\Gamma/2)F}
\right]
\label{eq:phcr_sil_condition}
\end{equation}

Here, $K_0$ is the SIL feedback coefficient determined by the laser output-coupling rate, the resonator--waveguide coupling rate, the resonator linewidth, and the laser linewidth-enhancement factor~\cite{Shen2020}
and can therefore be held constant even when the mode splitting $\Gamma$ is varied. The parameter $\bar\psi$ is the effective feedback phase. The feedback-induced frequency pulling depends on the amplitude and phase of the BW pump field, and $\zeta$ must be determined self-consistently with the intracavity fields.

To examine the stationary resonance structure, we first consider CW solutions that are homogeneous along the ring. Under this assumption, all sideband modes vanish, $\tilde a_{\mu\neq0}=\tilde b_{\mu\neq0}=0$. Setting $\partial_\tau\tilde a_0=\partial_\tau\tilde b_0=0$ and defining $a_0\equiv\tilde a_0$ and $b_0\equiv\tilde b_0$, the coupled LLEs reduce to
\begin{equation}
0
=
-(1+i\zeta)a_0
+i|a_0|^2a_0
+2i|b_0|^2a_0
+i\frac{\Gamma}{2}b_0
+F
\label{eq:CW_FW}
\end{equation}
\begin{equation}
0
=
-(1+i\zeta)b_0
+i|b_0|^2b_0
+2i|a_0|^2b_0
+i\frac{\Gamma}{2}a_0
\label{eq:CW_BW}
\end{equation}

The dispersion term vanishes because only the pumped modes ($\mu=0$) are present in the CW limit. The nonlinear terms give self- and cross-phase modulation of each CW component, while the coupling terms describe coherent energy exchange between the FW and BW pump modes. 
The CW fields must additionally satisfy the SIL condition given by Eq.~(3), with $\tilde b_0=b_0$ in the CW limit. The resulting self-consistent CW solutions describe the resonance branches and the nonlinear SIL tuning curve, which may be multivalued. These solutions are then used to evaluate the local stability of the SIL-locked states and the MI gain of non-pump modes. The subsequent comb formation and soliton dynamics are examined through direct time-domain simulations of the coupled LLEs.

\section{Results}

\subsection{Parametric investigation of soliton states}
\label{sec:results}

We first investigate the comb states accessible through turnkey operation of a self-injection-locked PhCR. To this end, we numerically integrate the coupled LLEs for different combinations of the free-running laser detuning $\xi$ and feedback phase $\bar\psi$, with the effective detuning $\zeta$ determined self-consistently by the SIL condition throughout the evolution. At each operating point, the FW and BW intracavity fields are initialized with noise and evolved independently while the external control parameters are held fixed. Thus, the turnkey simulations do not involve an externally imposed laser-frequency scan or parameter continuation between operating points. The final states are classified using the criteria described in \ref{app:classification}.

\begin{figure}[t]
\centering
\includegraphics[width=0.85\linewidth]{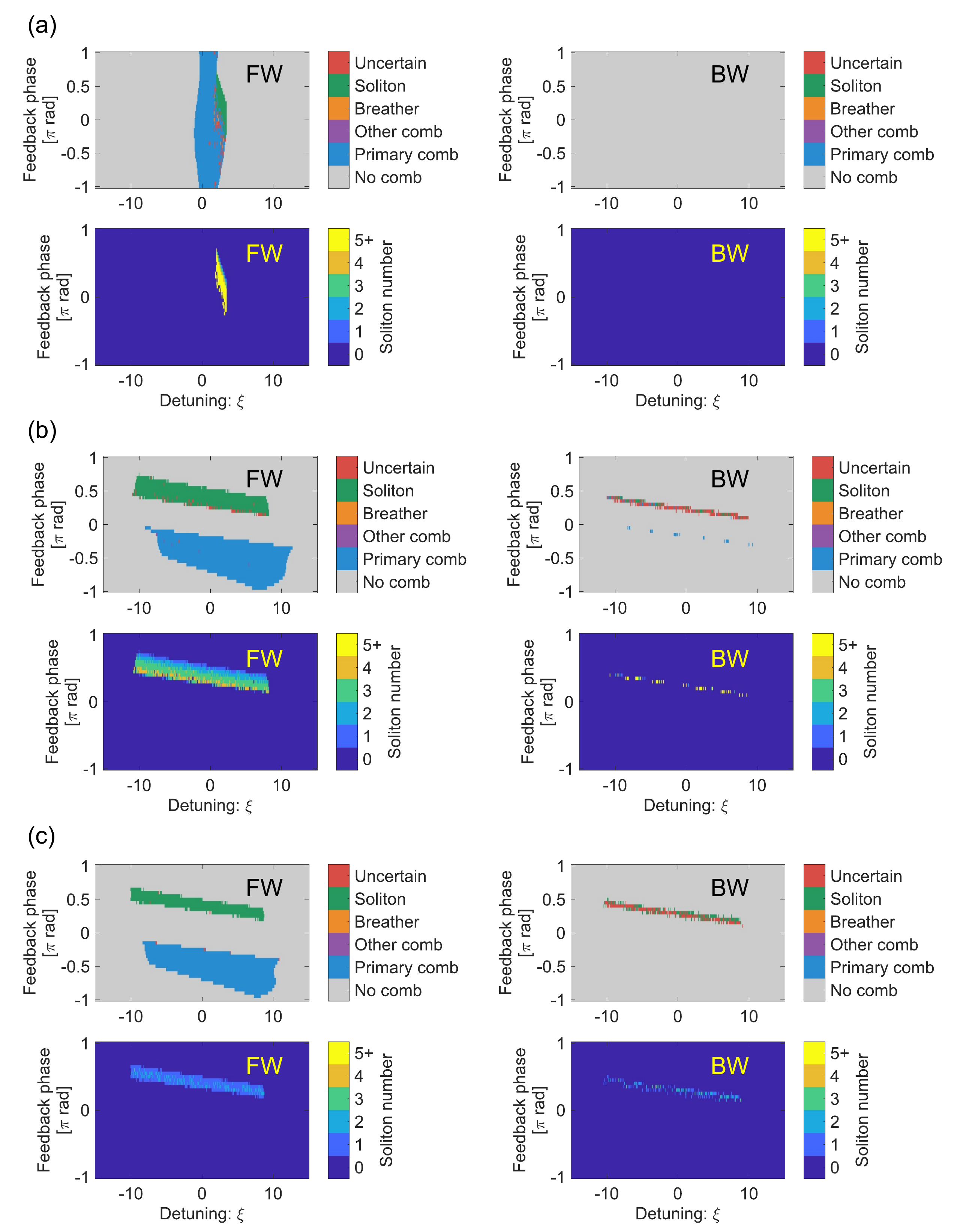}
\caption{Final-state maps obtained under turnkey operation as functions of the free-running detuning and feedback phase for (a) $\Gamma=0.1$, (b) $\Gamma=2.5$, and (c) $\Gamma=3$, calculated at $F^2=4$ and $K_0=45$. For each $\Gamma$, the upper panels show the comb-state classifications for the FW and BW fields, while the lower panels show the corresponding soliton numbers.}
\label{fig:turnkey_result}
\end{figure}

We fix the normalized pump power at $F^2=4$ to focus on stationary soliton operation, since higher pump powers promote breathing and chaotic comb states and reduce the parameter region yielding stationary solitons, as discussed in Section~\ref{sec:discussion}. The normalized second-order dispersion is set to $d_2=0.02$. At this pump power, the approximate analytical criteria for MI initiation and $m=1$ selection proposed by Ulanov et al.~\cite{Ulanov2024} cannot be satisfied simultaneously for any value of $\Gamma$. The feedback strength is fixed at $K_0=45$~\cite{nishimoto2026}, while the normalized pump-mode splitting is varied over $\Gamma=0.1$, $2.5$, $3$, and $4$. The weak-coupling case $\Gamma=0.1$ serves as a reference approximating a conventional ring resonator without appreciable engineered mode splitting, while retaining a finite BW field to provide optical feedback. Figure~\ref{fig:turnkey_result} shows the final-state maps in the $(\xi,\bar\psi)$ plane for both propagation directions. For each $\Gamma$, the upper panels show the comb-state classifications, whereas the lower panels show the corresponding soliton numbers. At $\Gamma=0.1$, FW solitons are obtained only within a relatively narrow parameter region, with multi-soliton states accounting for most of the soliton outcomes. The BW field remains predominantly in the no-comb state, consistent with weak FW--BW coupling. As $\Gamma$ increases to $2.5$ and $3$, FW soliton generation becomes accessible over relatively broad ranges of $\xi$ at selected feedback phases. 
At the same time, states with larger soliton numbers are progressively suppressed, shifting the soliton-number distribution toward smaller values, with single-soliton states dominating at $\Gamma=3$. Comb generation also appears over a limited region in the BW field. At $\Gamma=4$ (not shown in Fig.~\ref{fig:turnkey_result}), no soliton states are obtained in either propagation direction within the investigated parameter range, although primary-comb states originating from the upper CW branch of the upper resonance remain accessible in the FW field. Since BW comb generation is limited at $F^2=4$, only the FW results are presented in the following figures, although both fields are retained in the coupled-LLE simulations and stability analyses.

\begin{figure}[tbp]
\centering
\includegraphics[width=\linewidth]{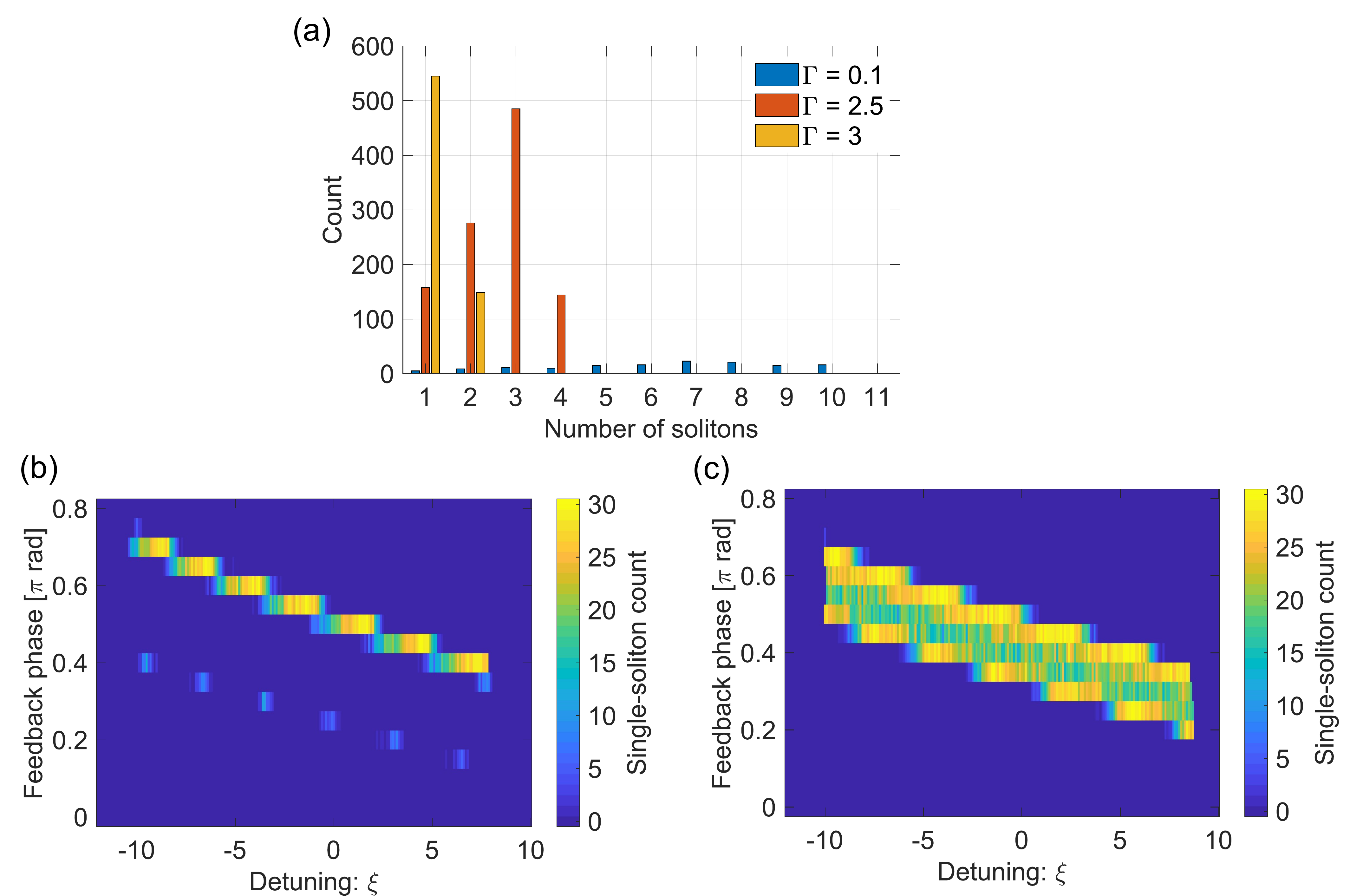}
\caption{Statistics of FW soliton states obtained under turnkey operation. (a) Number of sampled operating points yielding each FW soliton number $N_{\mathrm{FW}}$ for $\Gamma=0.1$, $2.5$, and $3$. (b,c) Number of FW single-soliton outcomes among 30 independent turnkey simulations at each combination of free-running detuning and feedback phase for (b) $\Gamma=2.5$ and (c) $\Gamma=3$.}
\label{fig:soliton_stat}
\end{figure}

To quantify these trends, Fig.~\ref{fig:soliton_stat}(a) shows the distribution of FW soliton numbers $N_{\mathrm{FW}}$ across the sampled operating points. The total number of operating points yielding FW solitons increases from 142 at $\Gamma=0.1$ to 1063 at $\Gamma=2.5$, then decreases to 695 at $\Gamma=3$. As $\Gamma$ increases, the distribution shifts toward lower soliton numbers. 
The most common soliton number decreases from seven at $\Gamma=0.1$ to three at $\Gamma=2.5$ and one at $\Gamma=3$. The number of single-soliton operating points increases from 5 at $\Gamma=0.1$ to 158 at $\Gamma=2.5$ and 545 at $\Gamma=3$, corresponding to approximately $3.5\%$, $14.9\%$, and $78.4\%$ of the FW soliton operating points, respectively. Thus, increasing $\Gamma$ from 2.5 to 3 reduces the overall number of FW soliton operating points but substantially enhances single-soliton selectivity.
We further examine the repeatability of turnkey single-soliton generation at $\Gamma=2.5$ and $\Gamma=3$. Figures~\ref{fig:soliton_stat}(b) and (c) show the number of single-soliton outcomes among 30 independent turnkey simulations at each $(\xi,\bar\psi)$ pair. At $\Gamma=2.5$, high success counts are limited to a small number of parameter combinations. In contrast, at $\Gamma=3$, high success counts extend over a substantially broader range of free-running detuning and feedback phase, with many operating points yielding single solitons in all 30 simulations. Therefore, despite the smaller total number of FW soliton operating points, $\Gamma=3$ provides a broader parameter region for highly repeatable turnkey single-soliton generation. The physical origin of these trends is examined in the following subsection through the CW resonance structure, MI gain, and nonlinear SIL tuning curve.

\subsection{Prediction of comb and soliton formation based on MI and SIL}
\label{subsec:mi_sil_stability}

\begin{figure}[b]
\centering
\includegraphics[width=\linewidth]{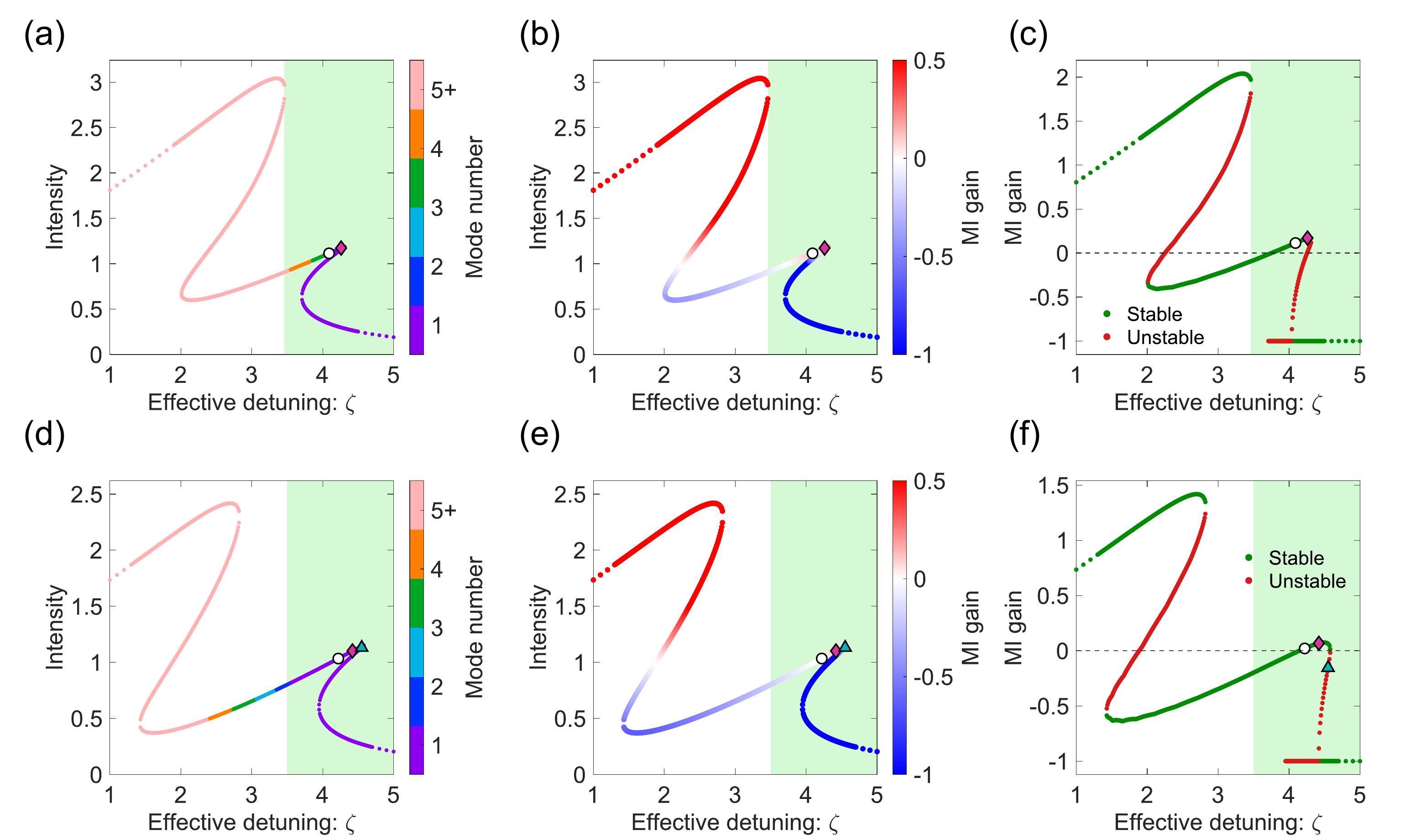}
\caption{
Modulation-instability characteristics of the FW CW steady-state solutions for $\Gamma=2.5$ (a--c) and $\Gamma=3$ (d--f). Panels (a,d) and (b,e) show the intracavity CW intensities as functions of the effective detuning $\zeta$, colored by the dominant sideband mode number and maximum MI gain, respectively. Panels (c,f) show the maximum MI gain as a function of $\zeta$, with green and red curves indicating stable and unstable CW branches, respectively. The light-green shaded regions indicate the estimated FW soliton existence ranges. 
The open black circles indicate the SIL-stable CW operating points on the lower resonance following the transition from the upper resonance induced by the loss of SIL stability, as illustrated in Fig.~\ref{fig:locking_curve}.
 The magenta diamonds and cyan triangles mark the operating points also indicated in Fig.~\ref{fig:locking_curve} and examined through time-domain simulations in Figs.~\ref{fig:comb_power}(c)--(e).
}
\label{fig:MI_gain}
\end{figure}

To understand the mechanisms governing turnkey comb generation and the resulting soliton states, we investigate the modulation instability of the CW solutions and the stability of the SIL-locked states. We first evaluate the MI gain through a linear stability analysis of the CW solutions, following the coupled-LLE approach developed in our previous work~\cite{Sakaue2026}. For each CW solution, we calculate the maximum MI gain from the real parts of the sideband eigenvalues ($m\geq1$). A positive MI gain indicates that the CW state is unstable to sideband perturbations, allowing comb generation to be initiated.

Figure~\ref{fig:MI_gain} shows the MI characteristics of the FW field along the CW resonance branches for $\Gamma=2.5$ and $\Gamma=3$. At $\Gamma=2.5$, the MI gain becomes positive at $\zeta\simeq3.76$ on the upper CW branch of the lower resonance. The positive gain persists up to the turning point at $\zeta\simeq4.31$ and extends into a portion of the intermediate CW branch down to $\zeta\simeq4.26$. At $\Gamma=3$, the MI gain becomes positive at $\zeta\simeq4.15$ and remains positive up to the turning point at $\zeta\simeq4.58$. Thus, increasing $\Gamma$ shifts the MI-positive region associated with the lower resonance toward larger effective detuning $\zeta$.
The mode number corresponding to the maximum MI gain also tends to decrease with increasing $\zeta$. At $\Gamma=2.5$, the dominant mode changes from $m=4$ at $\zeta\simeq3.86$ to $m=3$, $2$, and $1$ at $\zeta\simeq4.04$, $4.14$, and $4.19$, respectively. At $\Gamma=3$, the corresponding mode numbers are $m=4$, $3$, $2$, and $1$ at $\zeta\simeq2.74$, $3.07$, $3.35$, and $4.55$, respectively. These results indicate a tendency toward lower-order dominant modes at larger effective detunings. 
Both the shift of the MI-positive region and the change in the dominant MI mode are related to the resonance structure modified by mode splitting. In the linear limit, the lower and upper resonances are located approximately at $\zeta=+\Gamma/2$ and $\zeta=-\Gamma/2$, respectively. Increasing $\Gamma$ therefore shifts the lower resonance toward larger positive detuning, consistent with the observed shift of the MI-positive region. The dominant MI mode is further governed by the four-wave-mixing phase-matching condition. For anomalous dispersion ($d_2>0$), the phase-matched mode number can be approximated as
\begin{equation}
m_{\mathrm{pm}}^2 \simeq
\frac{2}{d_2}
\left[2\left(|a_0|^2+|b_0|^2\right)-\zeta\right],
\label{eq:phcr_sil_mi_peak_mode}
\end{equation}
when the right-hand side is positive. As $\zeta$ increases along a CW branch, the phase-matched mode number decreases if the increase in $\zeta$ exceeds that of the nonlinear shift $2(|a_0|^2+|b_0|^2)$. 
The estimated FW soliton existence ranges, calculated using the effective detuning and pump parameters introduced in our previous work~\cite{Sakaue2026}, are indicated by the light-green shaded regions in Fig.~\ref{fig:MI_gain}. For the parameters considered here, these ranges encompass all detuning regions where the FW MI gain is non-negative.

The MI gain in Fig.~\ref{fig:MI_gain} is evaluated as a function of the effective laser--cavity detuning $\zeta$. Under SIL, however, the externally controlled parameter is the free-running detuning $\xi$, while $\zeta$ is determined self-consistently through optical feedback. Consequently, a positive MI gain at a given $\zeta$ does not necessarily imply that the corresponding operating point is accessible under turnkey operation at a prescribed $\xi$ and $\bar\psi$. To determine the possible operating points, we calculate the nonlinear SIL tuning curve by solving Eqs.~\eqref{eq:CW_FW}, \eqref{eq:CW_BW}, and \eqref{eq:phcr_sil_condition} self-consistently. We further examine the local stability of the SIL-locked states against perturbations of the pumped modes, as described in \ref{app:sil_stability}.

\begin{figure}[tbp]
\centering
\includegraphics[width=\linewidth]{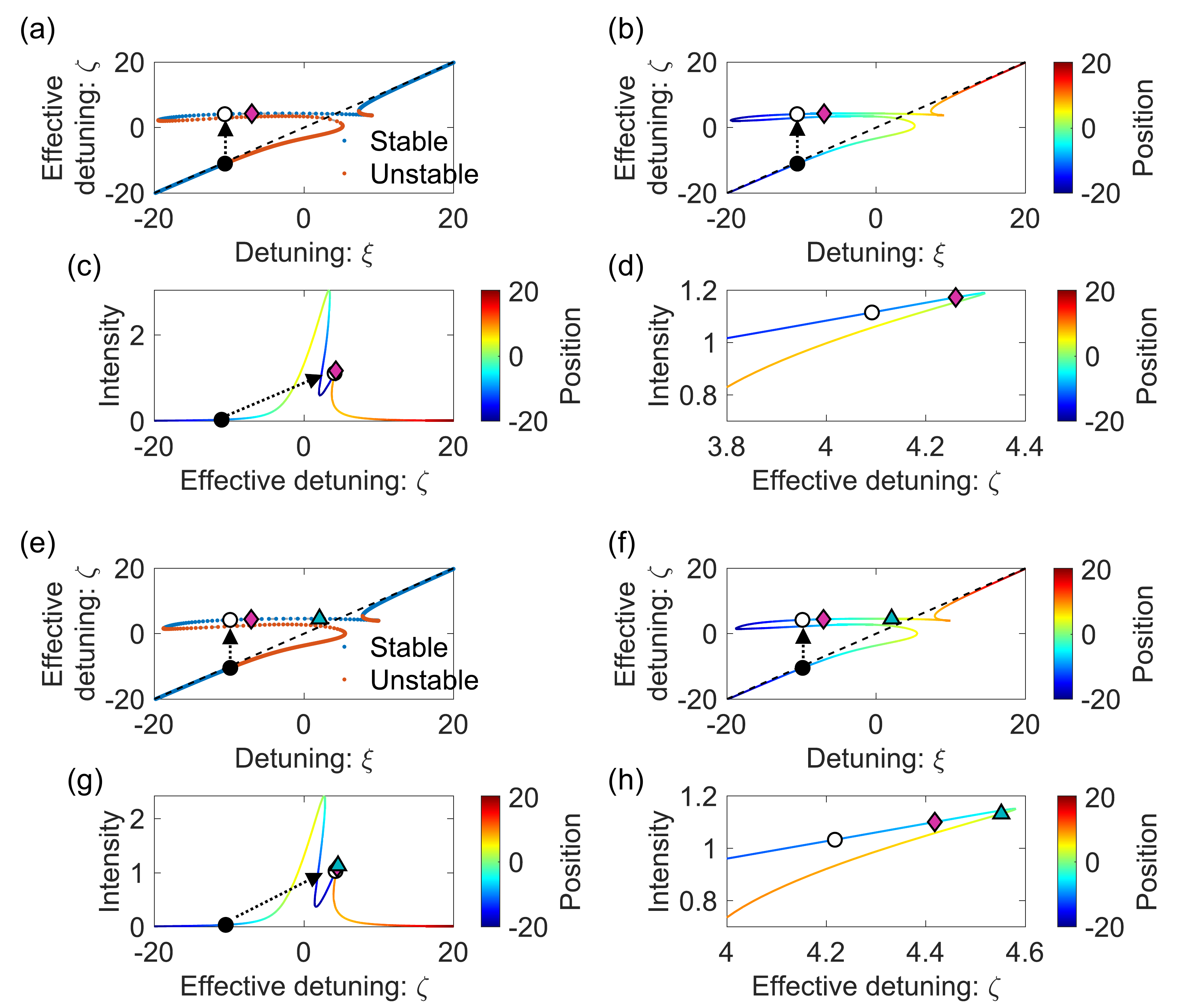}
\caption{
Nonlinear SIL tuning curves and CW steady-state solutions for $\Gamma=2.5$ (a--d) and $\Gamma=3$ (e--h), calculated at $\bar\psi=0.5\pi$. Panels (a,e) show the nonlinear SIL tuning curves, with blue and red indicating SIL-stable and SIL-unstable states, respectively. Panels (b,f) show the same curves, colored to indicate the corresponding positions on the CW resonance branches. Panels (c,g) show the FW intracavity CW intensities as functions of the effective detuning $\zeta$, while panels (d,h) provide enlarged views of the upper CW branch of the lower resonance. The black dashed lines in (a,b,e,f) represent $\zeta=\xi$. The filled black circles indicate the points where the upper CW branch of the upper resonance loses SIL stability, while the open circles indicate the corresponding stable operating points on the upper CW branch of the lower resonance. The dotted arrows illustrate the possible transitions between these branches. The magenta diamonds indicate the operating points at $\xi=-7$ for both values of $\Gamma$, while the cyan triangles indicate the operating point at $\xi=2$ for $\Gamma=3$. These colored markers correspond to those in Fig.~\ref{fig:MI_gain} and the time-domain simulations in Figs.~\ref{fig:comb_power}(c)--(e).
}
\label{fig:locking_curve}
\end{figure}

Figure~\ref{fig:locking_curve} shows the CW steady-state solutions and nonlinear SIL tuning curves for $\Gamma=2.5$ and $\Gamma=3$ at $\bar\psi=0.5\pi$. For both values of $\Gamma$, the upper CW branch of the upper resonance loses SIL stability as $\xi$ increases along the calculated tuning curves, at $\xi\simeq-10.55$ for $\Gamma=2.5$ and $\xi\simeq-9.84$ for $\Gamma=3$. At these respective values of $\xi$, SIL-stable solutions exist on the upper CW branch of the lower resonance at $\zeta\simeq4.09$ and $\zeta\simeq4.22$, respectively, as shown in Figs.~\ref{fig:locking_curve}(c), (d), (g), and (h). These results suggest that the loss of SIL stability can lead to a transition between the two branches, accompanied by abrupt changes in $\zeta$ and intracavity powers, as illustrated by the filled and open black circles in Fig.~\ref{fig:locking_curve}. The relatively large positive values of $\zeta$ reflect both mode splitting and Kerr-induced resonance shifts. On the upper CW branch of the lower resonance, $\zeta$ varies only weakly with $\xi$ owing to SIL feedback, allowing SIL-locked operating points to persist over a range of free-running detunings.

Combining the SIL and MI analyses clarifies the conditions for turnkey comb initiation and soliton formation. As shown in Figs.~\ref{fig:locking_curve}(d) and (h), SIL-stable CW solutions exist on the upper CW branch of the lower resonance at $\zeta\simeq4.09$ for $\Gamma=2.5$ and $\zeta\simeq4.22$ for $\Gamma=3$, respectively. At both operating points, the FW MI gain is positive, as shown in Figs.~\ref{fig:MI_gain}(c) and (f), indicating that sideband perturbations can grow and initiate comb generation if these CW states are reached. Furthermore, the MI-positive regions lie within the estimated FW soliton existence ranges indicated by the light-green shading in Figs.~\ref{fig:MI_gain}(c) and (f), allowing the subsequent nonlinear evolution to potentially produce soliton states.
The dominant MI modes also provide insight into soliton-number selectivity. At the SIL-stable CW operating points identified above, the dominant mode is $m=3$ for $\Gamma=2.5$ and $m=1$ for $\Gamma=3$, as indicated by Figs.~\ref{fig:MI_gain}(a) and (d). Higher-order dominant modes favor initial intensity modulations with multiple maxima, which may promote multi-soliton formation, whereas $m=1$ favors single-soliton formation. This difference is consistent with the enhanced single-soliton selectivity at $\Gamma=3$ observed in Figs.~\ref{fig:turnkey_result} and \ref{fig:soliton_stat}. Nevertheless, the dominant MI mode does not uniquely determine the final soliton number, which also depends on the subsequent nonlinear evolution, including feedback-induced changes in the effective detuning. These dynamics are examined in the following subsection through direct time-domain simulations.

\subsection{Numerical verification of turnkey soliton generation}
\label{subsec:time_evoution}

\begin{figure}[tbp]
    \centering
    \includegraphics[width=0.65\linewidth]{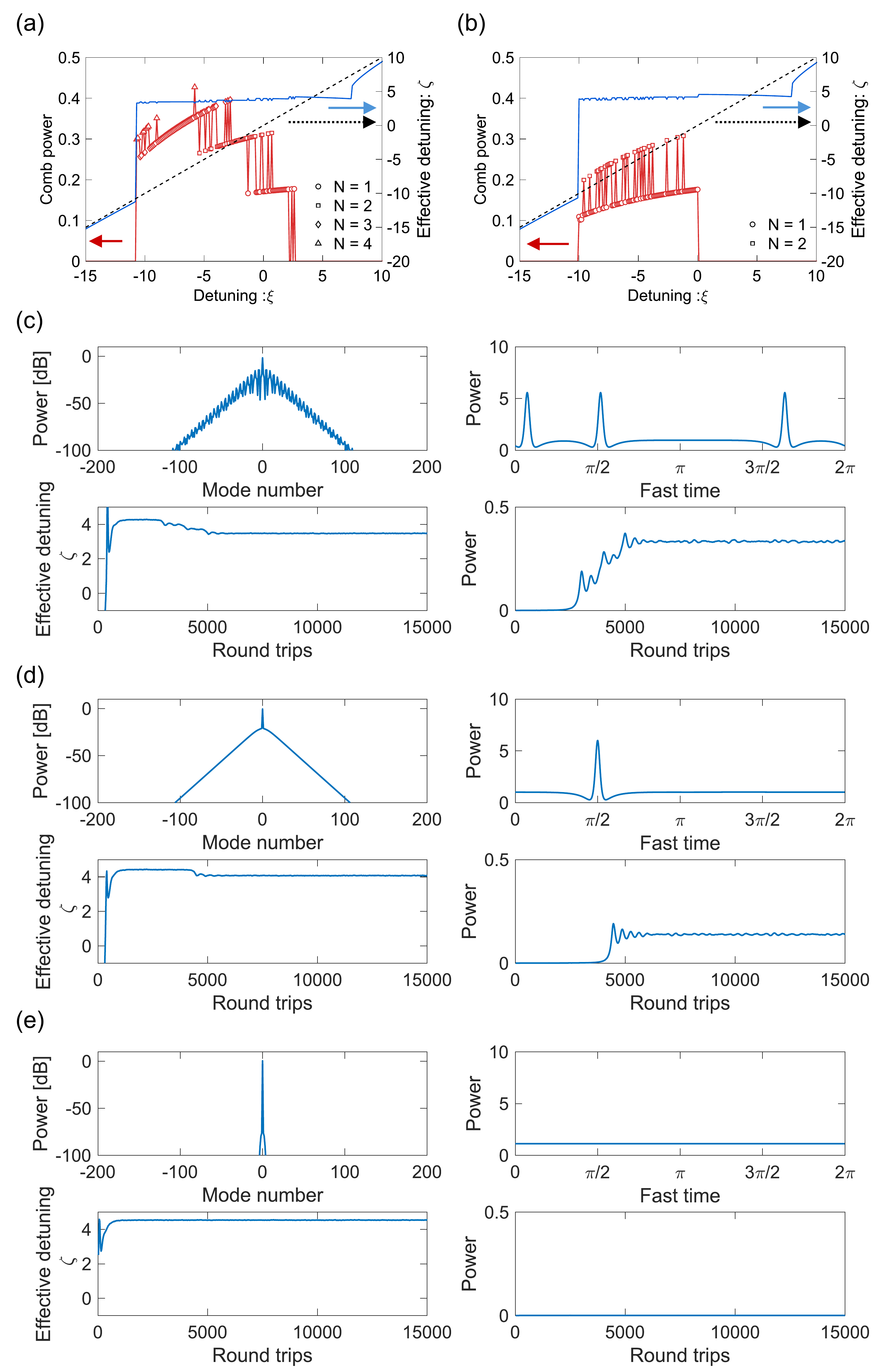}
\caption{Turnkey comb generation and temporal evolution of the intracavity fields. (a,b) Final FW comb power (red) and effective laser--cavity detuning $\zeta$ (blue) as functions of the free-running detuning $\xi$ for (a) $\Gamma=2.5$ and (b) $\Gamma=3$. The symbols indicate the number of FW solitons in the final states, and the black dashed lines represent $\zeta=\xi$. (c--e) Temporal evolution and final intracavity states for (c) $\Gamma=2.5$, $\xi=-7$, (d) $\Gamma=3$, $\xi=-7$, and (e) $\Gamma=3$, $\xi=2$. The upper panels show the final FW optical spectra (left) and intracavity intensity profiles (right), while the lower panels show the evolution of the effective detuning $\zeta$ (left) and FW comb power (right) as functions of the number of round trips. The feedback phase is fixed at $\bar\psi=0.5\pi$.}
    \label{fig:comb_power}
\end{figure}

To verify the predictions obtained from the MI and SIL analyses, we numerically integrate the coupled LLEs under turnkey operation and examine the resulting comb states. We focus on $\Gamma=2.5$ and $\Gamma=3$ at a fixed feedback phase of $\bar\psi=0.5\pi$. Figures~\ref{fig:comb_power}(a) and (b) show the final FW comb power and effective detuning $\zeta$ as functions of the free-running detuning $\xi$. The FW comb power is defined by excluding the pumped mode, $P_{\mathrm{comb,FW}}=\sum_{\mu\neq0}|\tilde a_\mu|^2$.

Figure~\ref{fig:comb_power}(a) shows that, at $\Gamma=2.5$, the effective detuning $\zeta$ changes abruptly from approximately $-11.22$ to $3.30$ at $\xi=-10.7$, accompanied by the onset of FW comb generation. FW soliton states are observed up to $\xi=2.6$, with isolated single-soliton states appearing between $\xi=-1.3$ and $2.6$. Similarly, Fig.~\ref{fig:comb_power}(b) shows an abrupt change in $\zeta$ from approximately $-10.68$ to $3.90$ at $\xi=-10.0$ for $\Gamma=3$, where FW comb generation begins. Soliton states are obtained over the range $-10.0\leq\xi\leq0.0$. The resulting soliton-number distributions differ markedly between the two cases. At $\Gamma=2.5$, 103 of the 130 FW soliton operating points yield multi-soliton states, whereas at $\Gamma=3$, 75 of the 101 FW soliton operating points yield single solitons.
The abrupt changes in $\zeta$ occur near the SIL stability boundaries identified in Fig.~\ref{fig:locking_curve}, at $\xi\simeq-10.55$ for $\Gamma=2.5$ and $\xi\simeq-9.84$ for $\Gamma=3$. This agreement is consistent with a transition from the upper CW branch of the upper resonance to the upper CW branch of the lower resonance following the loss of SIL stability. At the corresponding SIL-stable CW operating points, the maximum FW MI gains are approximately $0.11$ at $m=3$ for $\Gamma=2.5$ ($\zeta\simeq4.08$) and $0.017$ at $m=1$ for $\Gamma=3$ ($\zeta\simeq4.21$), as shown in Figs.~\ref{fig:MI_gain}(c) and (f). The positive MI gains are consistent with the onset of FW comb generation immediately after the abrupt changes in $\zeta$. The dominant MI modes also suggest different initial spatial scales of the growing perturbations, consistent with the greater occurrence of multi-soliton states at $\Gamma=2.5$ and single-soliton states at $\Gamma=3$. Notably, the effective detunings of the final comb states, $\zeta\simeq3.30$ and $3.90$, differ from those of the initial CW operating points because of the subsequent nonlinear evolution.

\begin{figure}[h]
    \centering
    \includegraphics[width=0.7\linewidth]{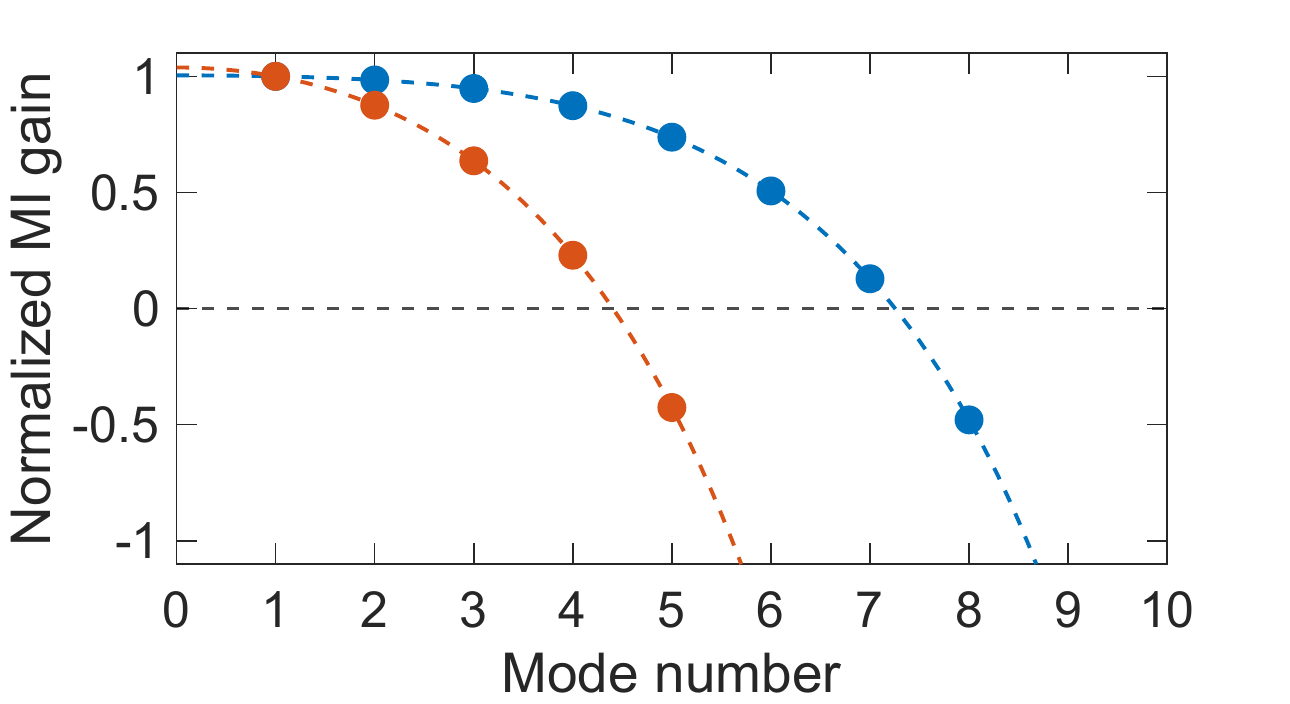}

\caption{FW MI gain, normalized to the gain at mode number $m=1$, as a function of sideband mode number for $\Gamma=2.5$ (blue) and $\Gamma=3$ (red), evaluated at the SIL-locked CW operating points corresponding to $\xi=-7$ and $\bar\psi=0.5\pi$.}

    \label{fig:MI_gain_spe}
\end{figure}

To further examine the dynamics of turnkey comb generation, we investigate the temporal evolution of the effective detuning $\zeta$ and FW comb power at representative operating points. Figures~\ref{fig:comb_power}(c)--(e) show the temporal evolution and final intracavity states for $(\Gamma,\xi)=(2.5,-7)$, $(3,-7)$, and $(3,2)$, respectively. 
For $\Gamma=2.5$ and $\xi=-7$ (Fig.~\ref{fig:comb_power}(c)), corresponding to the magenta diamonds in Figs.~\ref{fig:MI_gain}(a)--(c) and \ref{fig:locking_curve}(a)--(d), the system reaches a SIL-locked CW state after approximately 1300 round trips, with $\zeta$ increasing from near zero to approximately $4.27$, in agreement with the CW operating point predicted by the nonlinear SIL tuning curve in Fig.~\ref{fig:locking_curve}(a).
At this detuning, the maximum FW MI gain is approximately $0.17$, with a dominant mode number of $m=1$. Consequently, the FW comb power gradually increases, and a single soliton is initially formed after approximately 3000 round trips. As the comb develops, the intracavity pump-mode powers decrease, reducing the Kerr-induced resonance shift. Since the laser frequency follows the shifted resonance through SIL, the effective detuning $\zeta$ also decreases. During this evolution, additional solitons form, and the system successively evolves into two- and three-soliton states before reaching a stationary three-soliton state. The decrease in $\zeta$ shifts the phase-matching condition toward higher-order MI modes, which may contribute to the formation of additional solitons. However, the final soliton number is determined by the nonlinear dynamics and cannot be predicted solely from the MI gain of the initial CW state. 
For $\Gamma=3$ and $\xi=-7$ (Fig.~\ref{fig:comb_power}(d)), corresponding to the magenta diamonds in Figs.~\ref{fig:MI_gain}(d)--(f) and \ref{fig:locking_curve}(e)--(h), a similar process is observed.
Following the initial SIL locking, the FW comb gradually develops and a single soliton is formed. In contrast to the $\Gamma=2.5$ case, the single-soliton state remains stable without the formation of additional solitons, even though $\zeta$ decreases slightly during comb development. This difference can be understood from the MI-gain spectra shown in Fig.~\ref{fig:MI_gain_spe}. In both cases, the maximum MI gain occurs at $m=1$. However, for $\Gamma=2.5$, the MI gain is positive for seven mode numbers, whereas for $\Gamma=3$, it is positive for only four mode numbers and decays more steeply with increasing mode number. Thus, at $\Gamma=3$, the MI gain is more strongly concentrated around the lowest-order mode, favoring single-soliton formation. 
Finally, for $\Gamma=3$ and $\xi=2$ (Fig.~\ref{fig:comb_power}(e)), corresponding to the cyan triangles in Figs.~\ref{fig:MI_gain}(d)--(f) and \ref{fig:locking_curve}(e)--(h), SIL locking is established after approximately 1300 round trips, with the effective detuning reaching $\zeta\simeq4.53$.
At this operating point, the maximum FW MI gain is approximately $-0.23$, indicating that the CW state is stable against sideband perturbations. Consequently, no FW comb is generated, and the system remains in the CW state despite successful SIL locking.

\section{Discussion and conclusion}
\label{sec:discussion}

The preceding analyses provide a unified understanding of turnkey soliton generation in a PhCR based on the CW steady-state solutions, MI gain, and nonlinear SIL tuning curve. The SIL analysis identifies the possible operating points and their local stability, while the MI analysis predicts the onset of comb generation and provides an indication of the resulting soliton number. These predictions are consistent with the direct time-domain simulations. We further investigate the influence of dispersion on turnkey soliton generation to examine whether the same physical interpretation applies under different dispersion conditions.

\begin{figure}[t]
    \centering
    \includegraphics[width=0.9\linewidth]{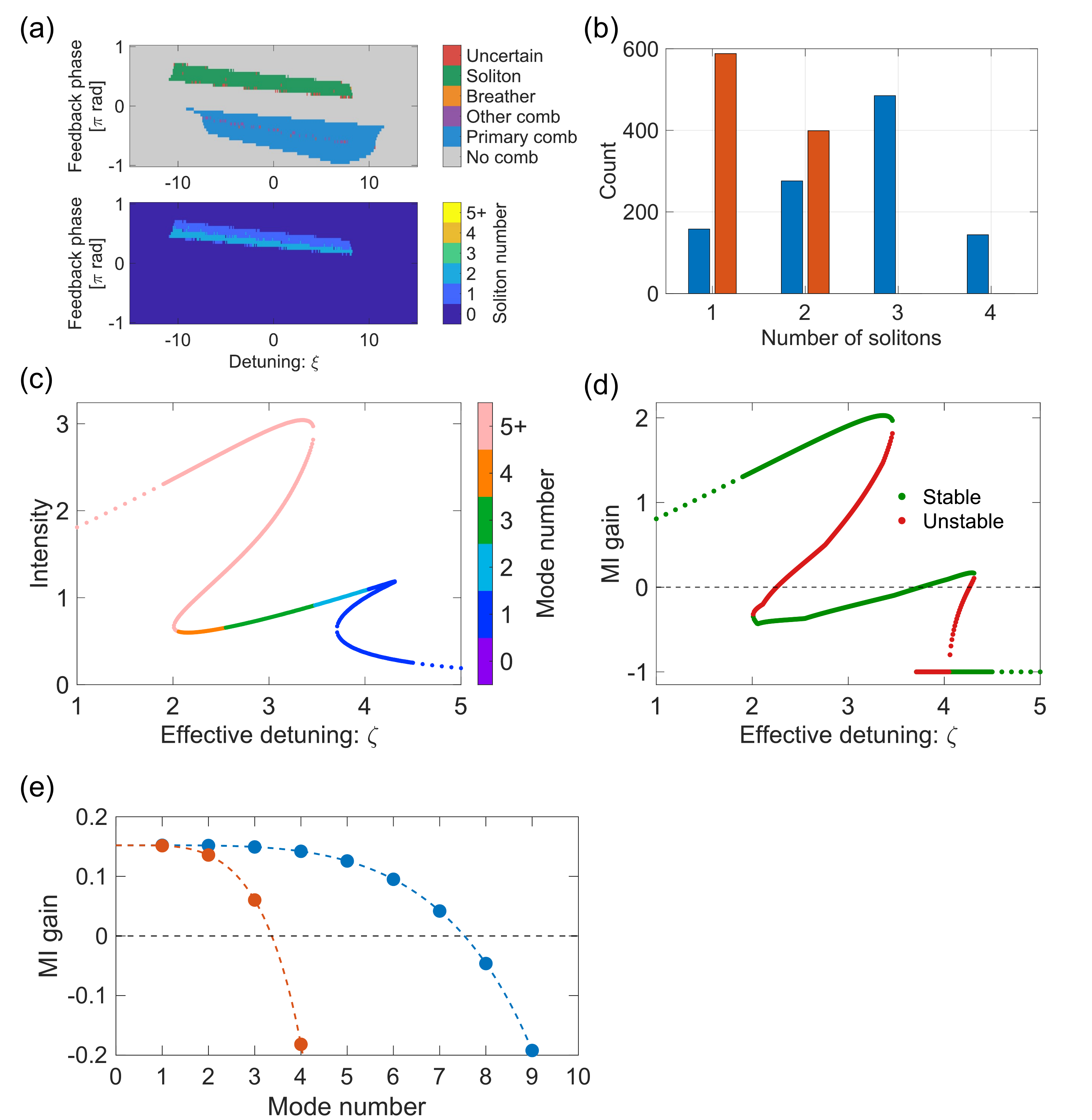}

\caption{Effect of increased dispersion on turnkey soliton generation at $\Gamma=2.5$, with the normalized second-order dispersion $d_2$ increased by a factor of five.
(a) Final-state map and soliton-number distribution for the FW field.
(b) Number of operating points yielding each FW soliton number $N_{\mathrm{FW}}$ for the original dispersion (blue) and fivefold dispersion (red).
(c) Dominant sideband mode number corresponding to the maximum FW MI gain along the CW resonance branches.
(d) Maximum FW MI gain as a function of the effective detuning $\zeta$.
(e) FW MI-gain spectra at $\zeta=4.2$ for the fivefold dispersion (red) and original dispersion (blue).
}
\label{fig:large_dispersion}
\end{figure}

Figure~\ref{fig:large_dispersion} shows the results obtained when the normalized second-order dispersion $d_2$ is increased by a factor of five, with the other parameters fixed at $\Gamma=2.5$. Figures~\ref{fig:large_dispersion}(a) and (b) show the final-state map and soliton-number distribution, respectively. The total number of operating points yielding FW solitons remains comparable between the two dispersion conditions, with 987 points for the fivefold dispersion and 1063 points for the original dispersion. However, the soliton-number distribution shifts toward lower soliton numbers as the dispersion increases, with single-soliton states becoming more prevalent.
Importantly, the nonlinear SIL tuning curve is independent of dispersion because it is determined from the CW steady-state equations. In contrast, dispersion directly affects the MI characteristics and subsequent nonlinear evolution, which can modify the final soliton states.
To understand the change in the soliton-number distribution, we examine the influence of dispersion on the MI characteristics. Comparing Fig.~\ref{fig:large_dispersion}(c) with Fig.~\ref{fig:MI_gain}(a), the mode number corresponding to the maximum FW MI gain is smaller at a given effective detuning $\zeta$ for the larger dispersion. This behavior is consistent with the phase-matching condition in Eq.~\eqref{eq:phcr_sil_mi_peak_mode}, which predicts a decrease in the phase-matched mode number with increasing $d_2$ for fixed intracavity powers and detuning. In contrast, the effective-detuning range over which the FW MI gain is positive remains approximately unchanged, as shown by comparing Fig.~\ref{fig:large_dispersion}(d) with Fig.~\ref{fig:MI_gain}(c).
To examine the MI characteristics in more detail, Fig.~\ref{fig:large_dispersion}(e) compares the FW MI-gain spectra at $\zeta=4.2$, where the maximum MI gain occurs at $m=1$ for both dispersion conditions. Although the peak MI gains are comparable, the gain decreases more rapidly with increasing mode number for the larger dispersion, resulting in fewer modes with positive MI gain. This narrower MI-gain spectrum suggests that fewer intensity maxima are favored during the initial stage of comb generation, which may contribute to the increased occurrence of low-soliton-number states. These results suggest that dispersion provides an additional degree of freedom for controlling soliton-number selectivity in self-injection-locked PhCRs.

\begin{figure}[t]
\centering
\includegraphics[width=0.85\linewidth]{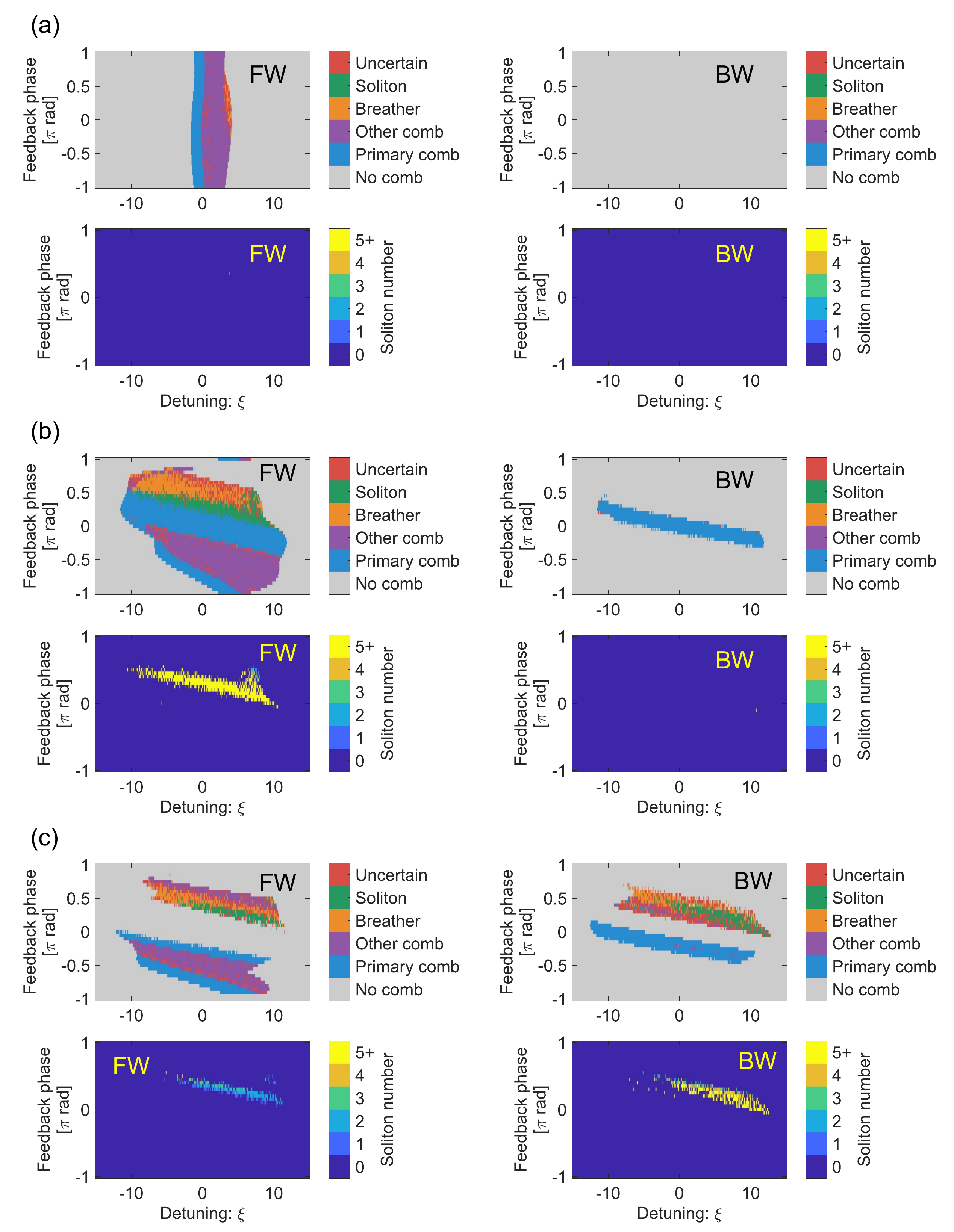}
\caption{
Final-state maps obtained under turnkey operation as functions of the free-running detuning and feedback phase for (a) $\Gamma=0.1$, (b) $\Gamma=2.5$, and (c) $\Gamma=5$, calculated at $F^2=7$ and $K_0=45$. For each $\Gamma$, the upper panels show the comb-state classifications for the FW and BW fields, while the lower panels show the corresponding soliton numbers.
}
\label{fig:turnkey_result_high_power}
\end{figure}

We next examine the influence of pump power on turnkey comb generation. Unlike dispersion, pump power modifies the CW steady-state solutions and thereby affects both the nonlinear SIL tuning curve and MI characteristics, influencing the accessible operating points and subsequent nonlinear evolution. Figure~\ref{fig:turnkey_result_high_power} shows the final-state maps obtained at a higher normalized pump power of $F^2=7$ for $\Gamma=0.1$, $2.5$, and $5$. Compared with the results at $F^2=4$ in Fig.~\ref{fig:turnkey_result}, the parameter regions yielding comb generation expand in both the FW and BW fields. However, breather and other nonstationary comb states, including chaotic combs, become more prevalent, while the regions yielding stationary solitons shrink.

In conclusion, our analysis reveals a possible pathway for turnkey comb initiation in self-injection-locked PhCRs. The loss of SIL stability on the upper CW branch of the upper resonance can enable a transition to the upper CW branch of the lower resonance, where positive MI gain supports comb formation. This mechanism is consistent with the abrupt detuning changes and comb onset observed in direct time-domain simulations. The final soliton number, however, cannot be determined solely from the dominant MI mode at the initial CW operating point. Even when the maximum MI gain occurs at $m=1$, the system can evolve into different soliton states, depending on the subsequent nonlinear evolution and feedback-induced detuning changes. The narrower MI-gain spectrum obtained at appropriate mode splitting favors single-soliton formation, consistent with the highly repeatable single-soliton generation observed in our simulations. We also find that increasing dispersion favors lower soliton numbers by suppressing higher-order unstable modes, whereas increasing pump power expands the comb-generation region but promotes nonstationary states at the expense of stationary solitons. These findings clarify the distinct roles of SIL stability, MI, and nonlinear evolution in turnkey soliton generation and provide guidance for controlling soliton-number selectivity in photonic-crystal ring resonators.

\appendix
\section{Classification of comb states}
\label{app:classification}

The final states obtained from the time-domain simulations are classified independently for the FW and BW fields based on their optical spectra, intracavity intensity profiles, and temporal dynamics. After the initial 20,000 round trips, the fields are propagated for an additional 18,000 round trips without added noise. The fields are sampled every 20 round trips, and the final 70\% of the recorded data is used for classification.

We first determine whether a comb is generated by evaluating the non-pump spectral power $P_{\mathrm{comb,FW}}=\sum_{\mu\neq0}|\tilde a_\mu|^2$ and $P_{\mathrm{comb,BW}}=\sum_{\mu\neq0}|\tilde b_\mu|^2$ for the FW and BW fields, respectively. Comb generation is identified when the corresponding non-pump spectral power exceeds $10^{-4}$, its fraction of the total spectral power exceeds $10^{-3}$, and at least two sideband lines exceed the prescribed spectral threshold. States without significant comb generation are classified as no-comb states.
For states exhibiting comb generation, we next examine the spatial morphology of the intracavity intensity profile. Primary-comb states are identified from regularly spaced intensity maxima and harmonically spaced spectral lines. Soliton-like states are identified by localized intensity peaks whose shapes are consistent with a $\mathrm{sech}^2$ profile. Specifically, the normalized shape error must be below 0.26, the intensity duty ratio must be below 0.25, and the pulse widths must satisfy the localization criteria. The number of pulses is determined by counting the localized intensity peaks. A state is considered to exhibit persistent soliton-like morphology when these characteristics are observed in at least 65\% of the sampled profiles.
The temporal dynamics are then evaluated from the variations in comb power, peak intensity, and detected peak count. A stationary state is identified when the relative standard deviations of the comb power and peak intensity are below 2.5\% and 3\%, respectively, and their relative drifts are below 6\%. For classification as either a stationary soliton or a breather, the detected peak count must differ by no more than one from its most frequently observed value in at least 90\% of the samples exhibiting localized profiles. Soliton-like states exhibiting sustained periodic oscillations are classified as breathers, with periodicity evaluated from spectral analysis and temporal autocorrelation.
Persistent comb states that do not satisfy the criteria for primary combs, stationary solitons, or breathers are classified as other combs. States exhibiting unresolved transients or insufficient evidence of convergence are classified as uncertain.

\section{Stability analysis of SIL states}
\label{app:sil_stability}

To evaluate the local stability of the SIL-locked CW solutions, we linearize the pump-mode equations around each steady state, including the feedback-induced change in the effective detuning $\zeta$. Writing $a_0=a_{\mathrm r}+ia_{\mathrm i}$ and $b_0=b_{\mathrm r}+ib_{\mathrm i}$, we define the real-valued state vector $\mathbf{x}=(a_{\mathrm r},a_{\mathrm i},b_{\mathrm r},b_{\mathrm i})^{\mathrm T}$. The pump-mode equations are expressed as
\begin{equation}
\frac{d\mathbf{x}}{d\tau}=\mathbf{G}(\mathbf{x},\zeta)=
\begin{pmatrix}
-a_{\mathrm r}+(\zeta-S_a)a_{\mathrm i}-(\Gamma/2)b_{\mathrm i}+F\\
-a_{\mathrm i}+(S_a-\zeta)a_{\mathrm r}+(\Gamma/2)b_{\mathrm r}\\
-b_{\mathrm r}+(\zeta-S_b)b_{\mathrm i}-(\Gamma/2)a_{\mathrm i}\\
-b_{\mathrm i}+(S_b-\zeta)b_{\mathrm r}+(\Gamma/2)a_{\mathrm r}
\end{pmatrix},
\end{equation}
where $S_a=|a_0|^2+2|b_0|^2$ and $S_b=|b_0|^2+2|a_0|^2$. The SIL condition in Eq.~(3) gives
\begin{equation}
\zeta(\mathbf{x})=\xi+\frac{K_0}{F}\left(b_{\mathrm i}\sin\bar\psi-b_{\mathrm r}\cos\bar\psi\right).
\end{equation}
For a CW steady state $\mathbf{x}_0$, a small perturbation $\delta\mathbf{x}$ satisfies
\begin{equation}
\frac{d}{d\tau}\delta\mathbf{x}=J\delta\mathbf{x},\qquad
J=\left[\left.\frac{\partial\mathbf{G}}{\partial\mathbf{x}}\right|_{\zeta}+\frac{\partial\mathbf{G}}{\partial\zeta}\left(\nabla_{\mathbf{x}}\zeta\right)^{\mathrm T}\right]_{\mathbf{x}=\mathbf{x}_0},
\end{equation}
where
\begin{equation}
\frac{\partial\mathbf{G}}{\partial\zeta}=
\begin{pmatrix}
a_{\mathrm i}\\
-a_{\mathrm r}\\
b_{\mathrm i}\\
-b_{\mathrm r}
\end{pmatrix},
\qquad
\nabla_{\mathbf{x}}\zeta=
\frac{K_0}{F}
\begin{pmatrix}
0\\
0\\
-\cos\bar\psi\\
\sin\bar\psi
\end{pmatrix}.
\end{equation}

The second term in $J$ accounts for changes in the effective detuning induced by perturbations of the BW pump field through SIL feedback. This contribution is not included in conventional CW stability analysis, where $\zeta$ is treated as a fixed parameter. The local stability is determined from the eigenvalues $\lambda_j$ of $J$. Defining $\Lambda_{\max}=\max_j\mathrm{Re}(\lambda_j)$, a CW state is classified as SIL-stable for $\Lambda_{\max}<0$ and SIL-unstable for $\Lambda_{\max}>0$.

\section*{Funding}
This work was supported by the Japan Society for the Promotion of Science (24K01386, 25H01887, 26K22740, and JPJS00420240022), Cabinet Office, Government of Japan (Subsidy for Reg. Univ. and Reg. Ind. Creation), and  JST FOREST Program (JPMJFR256Z). \\
\section*{Disclosures} The author declares no conflicts of interest.
\section*{Data availability} Data underlying the results presented in this paper are not publicly available at this time but may be obtained from the author upon reasonable request.

\bibliography{PhC_SILref}

\end{document}